\documentstyle[preprint,tighten,prl,aps]{revtex}
\begin{document}

\preprint{IFUP-TH 76/94}
\title{Running coupling constant and correlation length
from Wilson loops}
\author{Massimo Campostrini, Paolo Rossi, and Ettore Vicari}
\address{Dipartimento di Fisica dell'Universit\`a and I.N.F.N.,
I-56126 Pisa, Italy}
\maketitle

\begin{abstract}
We consider a definition of the QCD running coupling constant
$\alpha(\mu)$ related to Wilson loops of size $r{\times}t$ with
arbitrary fixed $t/r$.  The schemes defined by these couplings are
very close to the $\overline{\rm MS}$ scheme (i.e.\ the one-loop
perturbative correction to the coupling is small) for all values of
$t/r$; in the $t/r\to\infty$ limit, the ``$q\bar q$ force'' scheme is
recovered, where the coupling constant is related to the
quark-antiquark force.

We discuss the possibility of applying finite-size scaling techniques
to the Monte Carlo evaluation of $\alpha(\mu)$ up to very large
momentum scales.  We propose a definition of correlation length, also
related to Wilson loops, which should make such a computation
feasible.
\end{abstract}


One of the most important goals of a non-perturbative approach to QCD
such as the lattice is the determination of
$\Lambda_{\overline{\rm MS}}$ in units of a physical mass scale; this
determination requires a very precise measurement of the running
coupling constant in the $\overline{\rm MS}$ scheme
$\alpha_{\overline{\rm MS}}(\mu)$ in the high-momentum region, where
perturbation theory is reliable and accurate.  A determination of
$\alpha_{\overline{\rm MS}}(\mu{\sim}m_{Z_0})$ would be most welcome
from a phenomenological point of view.  Although considerable progress
has been achieved in the last few years, especially in quenched QCD
\cite{Luscher,Petronzio,Michael,Schilling,UKQCD,El-Khadra}, this task
turns out to be quite hard, due to the difficulty of performing
lattice calculations at large momentum scales.

In the present letter we will address the two problems of: choice of
an optimal $\alpha(\mu)$ for the lattice computation; determination of
$\alpha(\mu)$ up to the highest possible momentum scale.

The requirements for an optimal choice of running coupling constant
to be used in a lattice Monte Carlo simulation are:

1) fast and accurate lattice measurement;

2) proximity to $\alpha_{\overline{\rm MS}}$, to reduce systematic
errors due to neglected orders in the perturbative conversion to the
$\overline{\rm MS}$ scheme;

3) well-defined finite-size scaling properties.

We shall now present a new definition of $\alpha(\mu)$ addressing the
above points.

A natural definition of running coupling is derived from the static
quark-antiquark force $F(r)$ by the relationship
\begin{equation}
F(r) = - c_F {\alpha_{q{\bar{q}}}(1/r)\over r^2}\,,
\label{alphaqq}
\end{equation}
where $c_F=(N^2-1)/(2N)={4\over 3}$.  (We prefer to write
$\alpha_{q{\bar{q}}}$ with a momentum scale dependence, in analogy
with $\alpha_{\overline{\rm MS}}(\mu)$, rather then with a length
scale dependence, as often seen in the literature).  In perturbation
theory the ${q{\bar{q}}}$ and the ${\overline {\rm MS}}$ couplings are
related by a very small first-order coefficient~\cite{Billoire},
making the determination of $\alpha_{{\overline {\rm MS}}}$ from
$\alpha_{q{\bar{q}}}$ by a perturbative redefinition of the coupling
at large momentum very precise.

The force $F(r)$ between two static quarks separated by a distance $r$
can be evaluated from rectangular Wilson loops $W(r,t)$:
\begin{eqnarray}
&&F(r) = -{\partial V(r)\over \partial r}\,,\nonumber \\
&&V(r) = -\lim_{t\rightarrow\infty} \,
{1\over t}\ln W(r,t).
\label{force}
\end{eqnarray}
Recent developments in the calculation of $\alpha_{q{\bar{q}}}$ were
reported in Refs.~\cite{Schilling,UKQCD}.

On the lattice the implementation of this idea to extract
$\alpha_{q{\bar{q}}}(r)$ presents two major problems:

1) it requires a $t\rightarrow\infty$ limit procedure
in order to evaluate $V(r)$;

2) it requires control of the force from short to long distance in one
simulation, with $a\ll r\ll L$ holding for the whole range of physical
$r$ involved. This strongly limits the accessible values of $r$ since
the lattice size $L/a$ available on today's supercomputers is not very
large.

In this letter we extract $\alpha(\mu)$ from rectangular Wilson loops
$W(r,t)$ of size $r\times t$ with $x\equiv t/r$ fixed. This solves the
problem (1) above, and opens the road to the use of finite size
techniques in order to reach very small distances without the
necessity of very large lattices.  Moreover, the close connection with
the ${\overline {\rm MS}}$ scheme is retained, as we shall see from a
perturbative calculation.

In weak coupling,
\begin{equation}
\chi(r,t)\equiv {\partial^2 \ln W(r,t)\over \partial r \partial t}
 = - c_F A(x) {\alpha\over r^2}\left( 1 + O(\alpha)\right)\,,
\label{chi1}
\end{equation}
where $x=t/r$ and
\begin{equation}
A(x)  = {2\over \pi} \left[{\rm arctan} (x) \,+\, {1\over x} \,+\,
{{\rm arctan} (1/x)\over x^2}\right].
\label{ax}
\end{equation}
The renormalization properties of the Wilson loop
operator~\cite{Dotsenko,Aoyama} allow the definition of a running
coupling constant $\alpha_x(1/r)$ parametrized by $x=t/r$:
\begin{equation}
\chi(r,t) = - c_F A(x) {\alpha_x(1/r)\over r^2}\,.
\label{alphax}
\end{equation}
Without loss of generality, we will choose $r$ to be the shorter side
of the loop, and therefore $x\ge1$.
Since
\begin{equation}
\lim_{t\rightarrow\infty} \chi(r,t)= F(r),
\label{lf}
\end{equation}
we have
\begin{equation}
\lim_{x\rightarrow\infty} \alpha_x(1/r)= \alpha_{q{\bar{q}}}(1/r).
\label{lalpha}
\end{equation}

The perturbative expansion of $\chi(r,t)$ in dimensional
regularization can be obtained from the corresponding expansion of the
Wilson loop $W(r,t)$~\cite{VFC}.  In the ${\overline {\rm MS}}$
renormalization scheme
\begin{equation}
\chi(r,t) = - c_F A(x) {\alpha_{{\overline {\rm MS}}}(\mu)\over r^2}
\left[ 1 \,+\, b_0 \left( \ln (r\mu) +
R(x)\right) \alpha_{{\overline {\rm MS}}}(\mu)
\,+\, O\left(\alpha^2\right) \right],
\label{chip}
\end{equation}
where $b_0$ is the first coefficient of the $\beta$-function:
\begin{eqnarray}
&&\mu {\partial \over \partial \mu} \alpha_{{\overline {\rm MS}}} =
- b_0\alpha_{{\overline {\rm MS}}}^2 + O\bigl(\alpha^3\bigr),
\nonumber \\
&&b_0 = {1\over 4\pi}\left({22\over 3}N - {4\over 3}N_f\right),
\label{b0}
\end{eqnarray}
and $R(x)$ is a finite function of $x$:
\begin{eqnarray}
R(x) &=& \gamma_E - 1 + {1\over 4\pi b_0}
\left( {31\over 9}N-{10\over 9}N_f\right) \nonumber \\
&&\quad+\;{2\over\pi A(x)}\left[ I_1(x) + I_2(x)
+ {N\over 4\pi b_0}\left(I_3(x) + I_4(x) + I_5(x)\right)\right].
\label{Rx}
\end{eqnarray}
The functions $I_1,I_2,I_3,I_4,I_5$ are given in the Appendix.  They
all vanish in the large-$x$ limit where the perturbative expansion of
the static quark-antiquark force~\cite{Billoire} is recovered.

$R(x)$ determines the first order connection between the couplings
$\alpha_x(1/r)$ and $\alpha_{{\overline {\rm MS}}}(\mu)$:
\begin{eqnarray}
&&\alpha_{{\overline {\rm MS}}}(1/r) = \alpha_x(1/r) - b_0\,R(x)\,
\alpha^2_x(1/r) + O\left(\alpha^3_x(1/r)\right),\nonumber \\
&&\ln \left({\Lambda_x\over \Lambda_{{\overline {\rm MS}}}}
\right) = R(x),
\label{lambda}
\end{eqnarray}
where $\Lambda_x$ is the $x$-dependent $\Lambda$-parameter associated with
the $x$-schemes (\ref{alphax}).

In Table~\ref{ttt} we report $R(x)$ for $N_f=0$, 2, and 4, for various
values of $x$.  In quenched QCD ($N_f=0$), $R(x)$ turns out to be
independent of $N$; the values for $N_f\ne0$ in the Table are computed
for ${\rm SU}(3)$.  As anticipated above, the first order perturbative
coefficient relating $\alpha_x$ to $\alpha_{{\overline {\rm MS}}}$
remains small for all $x\geq 1$.

On the lattice we can define the quantity
\begin{equation}
\phi(r,x) = \chi_C\left({r\over a},{xr\over a}\right),
\end{equation}
where $\chi_C$ is the usual Creutz ratio
\begin{equation}
\chi_C(R,T) = -\ln {W(R,T)\,W(R{+}1,T{+}1)\over
W(R,T{+}1)\,W(R{+}1,T)}.
\label{crchi}
\end{equation}
$\phi(r,x)$ is a natural estimator of $\chi(r,xr)$.  By measuring
$\phi(r,x)$ at different scales $r$ keeping $x$ fixed, one may easily
extract the corresponding running coupling $\alpha_x(1/r)$.

We would like to conclude with some considerations concerning the
possibility of employing finite size techniques to reach very small
distances, without the need of very large lattices.  The power of this
kind of techniques was recently illustrated by the exploration of the
extremely small distance regime of two-dimensional spin models
\cite{Sokal}.

In the scaling region the following finite size scaling relations
holds
\begin{equation}
\xi_L(\beta)\simeq  f_\xi(L/\xi_L) \,\xi_\infty(\beta),
\label{xifss}
\end{equation}
where $\xi$ is a correlation length of the theory, and
\begin{equation}
\phi(\beta,r,x,L)\simeq
f_\phi(x,L/r,L/\xi_L) \,\phi(\beta,r,x,\infty),
\label{chifss}
\end{equation}
Notice that the existence of an additional scale $r$ in $\phi(r,x)$
causes an extra dependence on the ratio $L/r$ in the finite size
scaling function $f_\phi$.  Finite size scaling functions like $f_\xi$
and $f_\phi$ may be reconstructed by performing simulations at
relatively small lattices, by employing the scheme outlined in
Ref.~\cite{Sokal}.  In this context finite size methods should allow
us to keep $a\ll r$ even when $r$ is very small in physical unit.

A crucial point in the realization of this program is the definition
of a correlation length $\xi$ suitable for an accurate finite size
scaling study. Once found such a scale, the study of the finite size
scaling of $\phi$ should be relatively easy.  Finite size studies of
two-dimensional spin models\cite{Sokal,RV} suggest that a suitable
correlation length is one defined from the second moment of a
correlation function.  Such a correlation function should have
appropriate properties: it must renormalize multiplicatively with a
renormalization factor independent of the distance, fall down
exponentially at large distances, and of course be easily measurable
in Monte Carlo simulations.  We identified a possible candidate for
pure gauge theory, which is easily constructed from Wilson loops.  The
correlation function
\begin{equation}
Y(r,t)  = {W(r,t)\over W\bigl(\case{1}{2}r,\case{1}{2}t\bigr)^2}
\label{yrt}
\end{equation}
is renormalized by a constant factor, since the divergence associated
with the perimeter term of the Wilson loop operator~\cite{Aoyama}
cancels in the ratio; it is exponentially suppressed at large
distances, due to the area law of pure gauge theory; it is constructed
from Wilson loops, which are a standard ``easy'' observable of Monte
Carlo simulations.

{}From $Y(r,t)$
we can define a second moment type correlation length:
\begin{equation}
\xi^2 = {1\over 2}{\int_0^\infty {d} r \int_{r}^{\kappa r}{d} t \,
Y(r,t)\,rt \over \int_0^\infty {d} r
\int_{r}^{\kappa r}{d} t \, Y(r,t)}\,,
\label{xi}
\end{equation}
where $\kappa$ is a free parameter ($\kappa{>}1$), which can be chosen
to optimize the measurement.  Assuming an exact area law for the
Wilson loop, Eq.~(\ref{xi}) would give $\xi^2=1/\sigma$, where
$\sigma$ is the string tension.

The measurement of $\alpha_x(\mu)$ in the above-mentioned scheme up to
a large $\mu$ leads to a direct determination of the
$\beta$-function of the ${\rm SU}(3)$ lattice gauge theory and of the
adimensional quantity $\xi\Lambda_{\overline{\rm MS}}$.  This quantity
still needs to be converted to a more phenomenological scale, such as
$r_0$ \cite{Sommer}; but this is a rather minor problem, since it
involves only a measurement of $\xi/r_0$, which can be performed at
the values of $\beta$ of our choice.

The program presented here is not specific to pure gauge theories, and
both the definition of coupling constant and the finite-size scaling
relationships hold unchanged for full QCD, with the inclusion of
dynamical fermions.  The definition of correlation length has to be
changed; however a ``physical'' definition such as the inverse nucleon
mass is perfectly viable in this case.

\acknowledgments

It is a pleasure to thank G.~Paffuti and A.~Pelissetto
for useful and stimulating discussions.

\appendix
\section*{}

\begin{eqnarray}
&&I_1(x)  = {\pi\over2} A(x) - r{\partial\over \partial t}
\left[ f(x) + f(1/x)\right]
-r^2{\partial^2\over \partial r\partial t} \left[ f(1/x)\ln x\right],
\nonumber \\
&&I_2(x)  = -{\pi\ln 2\over2} \, A(x) - {1\over 2}r^2
{\partial^2\over \partial r\partial t} \left[ g(x)+g(1/x)\right],
\nonumber \\
&&I_3(x) = - r^2
{\partial^2\over \partial r\partial t} \left[ h(x)+h(1/x)\right],
\nonumber \\
&&I_4(x) = - r^2
{\partial^2\over \partial r\partial t} k(x),
\nonumber \\
&&I_5(x)  = l(x) + {l(1/x)\over x^2}\,,
\label{ifunc}
\end{eqnarray}
where $x=t/r$ and
\begin{eqnarray}
&&f(x) = x \,{\rm arctan}(x) - {1\over 2}\ln (1+x^2),
\nonumber \\
&&g(x) = x\int_0^x {d}y {\ln(1+y^2)\over 1+y^2}
- {1\over 4}\ln^2(1+x^2),
\nonumber \\
&&h(x) = f(x)\left( 6-4\ln x\right) - (1+x^2) {\rm arctan}^2 (x)
+2\int_0^x {d}y\left[ 2\ln y(x-y) + f(x-y)\right] {\rm arctan} (y),
\nonumber \\
&&k(x) = - 4 f(x) f(1/x),
\nonumber \\
&&l(x)={4\over\pi} \int_0^x{d}y \int_{-\infty}^\infty
{d} z_1 {d} z_2 \int_0^\infty {d} \rho \nonumber \\
&&\qquad\times\;
{\rho z_2 (y^2-x^2 + 2xz_1 - 2y z_1)\over
[z_1^2 + (z_2-1)^2 + \rho^2]
[(z_1-x)^2 + z_2^2 + \rho^2]^2 [(z_1-y)^2 + z_2^2 + \rho^2]^2}\,.
\label{i2func}
\end{eqnarray}



\begin{table}
\caption{We report $R(x) = \ln \Lambda_x/\Lambda_{{\overline {\rm MS}}}$
for $N_f=0,2,4$ and for some values of $x$.
}
\label{ttt}
\begin{tabular}{cr@{}lr@{}lr@{}l}
\multicolumn{1}{c}{$x$}&
\multicolumn{2}{c}{$N_f=0$}&
\multicolumn{2}{c}{$N_f=2$}&
\multicolumn{2}{c}{$N_f=4$}\\
\tableline \hline
1   & -0&.20758 & -0&.27875 & -0&.37270 \\
5/4 & -0&.11177 & -0&.18145 & -0&.27343 \\
4/3 & -0&.08975 & -0&.15852 & -0&.24930 \\
3/2 & -0&.05567 & -0&.12242 & -0&.21053 \\
5/3 & -0&.03131 & -0&.09603 & -0&.18146 \\
7/4 & -0&.02178 & -0&.08554 & -0&.16970 \\
2   & -0&.00069 & -0&.06189 & -0&.14266 \\
3   &  0&.03187 & -0&.02341 & -0&.09637 \\
4   &  0&.03979 & -0&.01317 & -0&.08308 \\
10  &  0&.04557 & -0&.00500 & -0&.07175 \\
$\infty$  &  0&.04691 & -0&.00324 & -0&.06945 \\
\end{tabular}
\end{table}

\end{document}